Relationships between six cultural scales and ten ageism dimensions: Correlation analysis using data from 31 countries

Keisuke Kokubun[1*]

[1]Graduate School of Management, Kyoto University, Kyoto, Japan
* kokubun.keisuke.6x@kyoto-u.jp

Abstract

As the aging of the world accelerates, clarifying the relationship between cultural differences and ageism is an urgent issue. Therefore, in this study, we conducted a correlation analysis between the six cultural scales of Hofstede et al. [1] and the 10 ageism scales calculated from data on 35,232 people from 31 countries included in the World Values Survey Wave 6 by Inglehart et al. [2]. The results of a partial correlation analysis controlling for economic and demographic factors showed that the cultural scales were correlated with ageism. This is the first study to show that diverse cultural scales are related to multiple dimensions of ageism.

1. Introduction

Ageism is the process of systematic stereotyping and discrimination against people because of their age [3]. Previous studies have found that ageism is related to the size of the aging population [4] and the rate of aging [5]. However, compared to these demographic discussions, there has been less discussion of cultural factors, and a recent systematic review of ageism across 199 papers showed that only eight studies considered institutional or cultural determinants [6]. Given that the world is made up of diverse cultures, research is valuable to gain insight into how cultural differences affect ageism.

As I will discuss later, recent studies have shown that certain cultures correlate with ageism using certain scales. However, to the author's knowledge, there seems to be no research that has comprehensively clarified this. At present, there is no consensus on what scale to use to measure ageism, and it has been pointed out that this may be one of the reasons why studies do not provide consistent results on the issue of whether ageism is stronger in the East or the West [7]. Therefore, in this study, using 10 items of ageism-related attitude data from the World Values Survey Wave 6 [2] and the values of six cultural scale indicators from Hofstede et al. [1], we show that diverse national cultures are associated with diverse ageism through partial correlation analysis that controls for economic and demographic factors.

2. Review of previous research and presentation of hypotheses

2.1. Hofstede's six cultural scales

The most widely used scales for measuring national culture are the six cultural scales by Hofstede et al. [1]. First, power distance (PDI) indicates the extent to which the less powerful members of an organization or institution accept that power is distributed unequally. A higher index indicates that hierarchies are clearly separated in society and are not questioned by its members. A lower index, on the other hand, means that people question authority and seek to distribute power equally [1]. Meanwhile, individualism (IDV) indicates the extent to which people in a society are integrated into groups. Individualistic societies often have loose ties that relate individuals only to their immediate family. They emphasize "I" and "we". At the other end of the spectrum, collectivist societies have tightly integrated relationships that bind extended family and others to in-groups. These in-groups are filled with unquestioning loyalty and support each other when conflicts arise with other in-groups [1].

Uncertainty avoidance (UAI) is a society's tolerance for ambiguity. In societies with high UAI, people avoid the unexpected, unknown, or unconventional. To do so, they adhere to strict codes of conduct, guidelines, and laws, and rely on absolute or unique truths. On the other hand, the lower the UAI, the more readily different thoughts and ideas are accepted. Such societies tend to be less regulated, more comfortable with ambiguity, and more liberal in their environments [1]. Masculinity (MAS) is the degree to which a society values achievement, its heroism and assertiveness, and its preference for material rewards for success. Societies with low UAI tend to value cooperation and humility, have compassion for the weak, and prefer quality of life [1].

Long-term orientation (LTO) links past connections to current and future actions and challenges. A high level of this indicator indicates that traditions are respected, maintained, and immutable [1]. Indulgence versus restraint (IVR) refers to the degree of freedom that social norms give to citizens. A society with high indulgence is one that considers the enjoyment of life to be a natural human desire and seeks to satisfy it freely. On the other hand, a society with high restraint is one that tries to control the satisfaction of desires and regulate them through strict social norms [1].

2.2. Hypothesis presentation regarding the relationship between cultural measures and ageism

This study shows that six factors, namely PDI, IDV, UAI, MAS, LTO, and IVR, are associated with ageism.

PDI: To the author's knowledge, there have been no studies clarifying the relationship between PDI and ageism. This may be because the two are conceptually close and therefore difficult to raise as subjects of analysis. According to Hofstede et al. [1], in societies with a high PDI, respect for parents and older relatives is a fundamental virtue that continues throughout life. Societies with a high PDI overlap geographically with Confucian societies such as China [8]. However, North and Fiske [5] found in a meta-analysis review of 37 papers that Eastern cultures have more negative attitudes toward the elderly than Western cultures. The reason for this is that Western societies with strong post-materialist values place as much importance on the welfare and dignity of the elderly as they do on individuals, and the positive effects of this outweigh the negative effects of the loss of the tradition of placing importance on the elderly [5,9]. These arguments are consistent with an earlier, paradoxical finding that the world's three highest elder suicide rates belong to South Korea, Taiwan, and China [10]. Thus, the following hypothesis is derived:

H1. The PDI is positively correlated with ageism.

IDV: A common belief is that Eastern cultures, with their strong collectivist traditions of filial piety, value older people more highly than Western cultures. However, the results of a meta-analysis reveal that ageism is stronger in collectivist cultures than in individualist cultures [5]. This is interpreted as being because collectivism, which emphasizes the differences between in-groups and out-groups, is more likely to produce negative stereotypes about different age groups, which may in turn fuel resentment toward older people who seek support and enjoy benefits without contributing to society [5]. Thus, the following hypothesis is derived:

H2. IDV is negatively correlated with ageism.

UAI: Societies with high uncertainty avoidance scores tend to be uncomfortable with unpredictability [1]. The experience of aging is largely unpredictable and uncontrollable, which may be a source of discomfort for people living in cultures that favor predictability [11, 12]. Consistent with these arguments, previous studies have consistently found that people from cultures with high uncertainty avoidance have more negative perceptions of aging [4, 11, 13]. For example, in a study by Ackerman and Chopik [11], people living in cultures with high uncertainty avoidance tended to have less warmth toward older people. Thus, the following hypothesis is derived:

H3. UAI is positively related to ageism.

MAS: Ng & Lim-Soh [14] conducted a study of 20 English-speaking countries and found that ageism in each country, assessed using a database of 8 billion words, correlates with MAS. MAS is related to ageism because a society that values competition and values the strong and successful easily sees older people, who are the opposite of competition, as weak [14]. Previous research has described, for example, how older men are alienated from younger men in their prime in social clubs in heavy industrial areas of England, where the economy is supported by male manual labor [15]. Relatedly, an analysis using large-scale data from the same World Values Survey Wave 6 [2] as this study showed that people who have a strong "market mentality" that emphasizes becoming rich and being successful in society are more likely to perceive the elderly as a burden on society. This is because people who are obsessed with money and success are more sensitive to the decline in their share due to the increase in the social burden of supporting the elderly [16]. Therefore, the following hypothesis is derived.

H4. MAS is positively correlated with ageism.

LTO: A study by Ackerman and Chopik [11] found that people living in countries with a higher long-term orientation (more emphasis on the future) tend to have higher prejudice against the elderly and less warmth toward the elderly. Similarly, the study by Ng and Lim-Soh [14] mentioned above shows that in addition to masculinity, ageism is more likely to be strengthened in societies with a long-term orientation. The common interpretation given by the authors of these studies is that, from a long-term perspective, investing in young people is expected to produce greater returns than investing in older people [11,14]. Thus, the following hypothesis is derived:

H5. LTO is positively related to ageism.

IVR: To the author's knowledge, there has been no research clarifying the relationship between IVR and ageism. According to Hofstede et al. [1], the world can be divided into ample and restrained societies. Ample societies are societies that seek to freely satisfy human desires related to savoring and enjoying life. On the other hand, restrained societies are societies that believe that strict social norms should suppress and limit the fulfillment of desires. Previous research has shown that ample societies tend to have lower mortality rates [1] and longer life expectancies [17]. The reason for this is thought to be that ample societies usually have a higher subjective sense of well-being and a positive view of happiness, which suppresses deaths from stress-related diseases such as cardiovascular disease (however, ample societies have a negative side in that people tend to consume more fast food and soft drinks, which makes them more likely to become obese). In other words, a fulfilling society where people enjoy life is a society where it is easy for the elderly to live. If the elderly are lively in a society, it is thought that discrimination and prejudice based on age are less likely to occur. Therefore, the following hypotheses are derived:

H6. IVR is negatively correlated with ageism.

3. Method

3.1. Sample and data collection

The data used were taken from Inglehart et al. [2] and Hofstede et al. [1]. The database of Inglehart et al. [2] was collected between 2010 and 2014 and includes 89,565 respondents living in 60 countries. From this, data from 35,232 people aged 16 to 59 years in 31 countries available from the database of Hofstede et al. [1] were used for the analysis. Table 1 shows the number of participants by country and survey year. The data of people aged 60 years and over were excluded mainly to minimize the influence of a certain "ownership" of people who are just about to become elderly "70 years and over" on the answer results. This decision was also partly because some countries with a small elderly population define elderly people as "60 years and over" in the questionnaire. We calculated the average values by country for the various ageism items listed below from Inglehart et al. [2] and calculated the correlation coefficients with the average values by country for the six cultural scales listed in Hofstede et al. [1].

Table 1. Number of participants by country and survey year.

| Country and survey year | N |
|---|---|
| Argentina 2012/13 | 815 |
| Australia 2012 | 1,089 |
| Brazil 2014 | 1,214 |
| Chile 2012 | 801 |
| China 2012/13 | 1,945 |
| Colombia 2012 | 1,294 |
| Estonia 2011 | 1,108 |
| Germany 2013 | 1,406 |
| Hong Kong 2014 | 788 |
| Japan 2010 | 1,571 |
| South Korea 2010 | 977 |
| Malaysia 2012 | 1,177 |
| Mexico 2012 | 1,793 |
| Morocco 2011 | 1,076 |
| Netherlands 2012 | 1,119 |
| New Zealand 2011/12 | 547 |
| Pakistan 2012 | 1,148 |
| Peru 2012 | 1,031 |
| Philippines 2012 | 1,001 |
| Poland 2012 | 689 |
| Romania 2012 | 1,105 |
| Russia 2011 | 1,972 |
| Singapore 2012 | 1,484 |
| Slovenia 2011 | 758 |
| Spain 2011 | 854 |
| Sweden 2011 | 852 |
| Thailand 2013 | 1,050 |
| Trinidad and Tobago 2010/11 | 728 |
| Turkey 2011 | 1,406 |
| United States 2011 | 1,683 |
| Uruguay 2011 | 751 |

3.2. Measures

Ageism was evaluated using the 10 variables shown in Table 2. Each variable consists of one item. As can be seen from the sentences, the items include those that ask about the general position and evaluation of older people from the perspective of others ("Social position: People in their 70s", "People over 70: are seen as friendly", "People over 70: are seen as competent", "People over 70: viewed with respect", "Older people are not respected much these days") and those that require subjective evaluation ("Is a 70-year old boss acceptable", "Older people get more than their fair share from the government", "Older people are a burden on society", "Companies that employ young people perform better than those that employ people of different ages", "Older people have too much political influence"). For this reason, it is difficult to create a reliable scale that can be evaluated by alpha coefficients by compressing the dimensions of these items through factor analysis [16], so in this study, we decided to use one item per scale variable for analysis.

Table 2. Ageism item names and scoring methods.

| Item | Scoring method |
|---|---|
| "Social position: People in their 70s" (n = 33,402). | The analysis in this study used scores from 1 to 10 given for the 10-point Likert scale options ranging from "extremely low" to "extremely high." |
| "People over 70: are seen as friendly" (n = 33,619), "People over 70: are seen as competent" (n = 33,421), "People over 70: viewed with respect" (n = 33,831). | The six-point Likert scale options, ranging from "Not at all likely to be viewed that way" to "Very likely to be viewed that way," were scored from 0 to 5 and used for the analysis in this study. |
| "Is a 70-year old boss acceptable" (n = 33,878). | The analysis in this study used scores from 1 to 10 given for the 10-point Likert scale options ranging from "completely unacceptable" to "completely acceptable." |
| "Older people are not respected much these days" (n = 34,203), "Older people get more than their fair share from the government" (n = 33,159), "Older people are a burden on society" (n = 33,885), "Companies that employ young people perform better than those that employ people of different ages" (n = 32,456), "Old people have too much political influence" (n = 31,904). | A four-point Likert scale ranging from "Strongly agree" to "Strongly disagree" was used in the analysis of this study with the scores from 1 to 4 given for each option reversed. |

The item names are taken from the variable names included in the database by Inglehart et al. [2] and are different from the questions in the questionnaire. In some countries with a small elderly population, the definition of elderly in the questionnaire is "60 years old or older." Therefore, the question text corresponding to the above item names is also changed to "60" instead of "70."

3.3. Control variables

There is evidence that the values of individuals are related to the economic level of a country [18,19]. Therefore, evaluations of the elderly have also changed in step with the global industrialization and lifestyle changes that have occurred over the past 200 years [14]. Therefore, in previous studies, economic indicators such as the level of modernization [16] and GDP per capita (Chen et al., 2023) are often used in analyzing the determinants of ageism. On the other hand, previous studies have shown that the aging rate, which is expressed by the growth rate of the population aged 65 and over [5,16] and the burden of supporting the elderly, such as the public pension rate [20], are a threat to resources and are related to the rise of ageism. In addition, the aging rate, which is expressed by the population ratio aged 65 and over, has also been shown to be related to ageism [4,16,20]. Therefore, in this study, GDP per capita, aging rate, and aging rate are used as covariates. Of these, GDP per capita is the natural logarithm of gross domestic product per capita converted to US dollars at constant prices. Meanwhile, the aging rate is the ratio of those aged 65 or over to the total population, and the aging rate indicates the change in the aging rate over the past 10 years. Both were obtained from the World Bank database [21], and unavailable data was supplemented with the United Nations database [22].

3.4. Analysis method

Correlation coefficients and partial correlation coefficients were calculated for the six cultural variables and the 10 ageism variables. The significance level was set at 5%. All statistical analyses have been performed using IBM SPSS Statistics Version 26 (IBM Corp., Armonk, NY, USA).

4. Analysis and findings

Table 3 shows the mean, standard deviation (SD), and correlation coefficient of each variable. The correlation coefficients below the diagonal are normal correlation coefficients, and those above the diagonal are partial correlation coefficients. Below, we will discuss the partial correlation coefficient results in order. First, PDI showed a significant positive correlation with "Older people are a burden on society" ($r = 0.393*$, $p = 0.039$), "Companies that employ young people perform better than those that employ people of different ages" ($r = 0.549**$, $p = 0.002$), and "Old people have too much political influence" ($r = 0.486**$, $p = 0.009$). On the other hand, IDV showed a significant negative correlation with "Old people have too much political influence" ($r = -0.400*$, $p = 0.035$). These results support H1 and H2. MAS did not show a significant correlation with any of the variables. Therefore, H3 is rejected. The UAI showed a significant positive correlation with "Older people are not respected much these days" ($r = 0.452*$, $p = 0.016$) and a significant negative correlation with "Older people get more than their fair share from the government" ($r = -0.608**$, $p = 0.001$). The former supports H4, while the latter contradicts H4. The lTOWVS showed a significant positive correlation with "Companies that employ young people perform better than those that employ people of different ages" (r = 0.450*, p = 0.016). This supports H5. Finally, IVR showed a significant positive correlation with "People over 70: are seen as friendly" (r = 0.424*, p = 0.024), and a significant negative correlation with "Older people are a burden on society" (r = -0.494**, p = 0.008) and "Companies that employ young people perform better than those that employ people of different ages" (r = -0.666***, p = 0.000). These support H6. Considering the large number of tests (60 times, i.e., 6 culture scales × 10 ageism scales), if we evaluate from the viewpoint of Bonferroni multiple comparison, the p-value for significance is 0.001 (= 0.05 ÷ 60), and only the two negative correlations, H4 UAI and "Older people get more than their fair share from the government" and H6 IVR and "Companies that employ young people perform better than those that employ people of different ages", survive as significant results. Also, the correlation coefficients in the lower left of Table 2 show a tendency roughly like the partial correlation coefficients in the upper right. Here, we would like to emphasize that PDI shows a significant positive correlation with "People over 70: viewed with respect" (r = 0.453*, p = 0.011), which contradicts the other ageism items. Figures 1 and 2 are scatter plots showing the two correlations that were significant in these multiple comparisons.

Table 3. Descriptive statistics and correlation coefficients

| | | Mean | SD | 1 | 2 | 3 | 4 | 5 | 6 | 7 | 8 | 9 | 10 | 11 | 12 | 13 | 14 | 15 | 16 |
|---|---|---|---|---|---|---|---|---|---|---|---|---|---|---|---|---|---|---|---|
| 1 | pdi | 61.710 | 19.683 | | -0.537** | 0.044 | 0.121 | 0.319 | -0.327 | -0.151 | -0.160 | -0.033 | 0.311 | -0.319 | 0.135 | 0.289 | 0.393* | 0.549** | 0.486** |
| 2 | idv | 40.870 | 23.332 | -0.639*** | | -0.027 | 0.191 | -0.202 | -0.280 | 0.313 | 0.086 | 0.150 | -0.015 | -0.042 | 0.024 | -0.356 | -0.043 | -0.351 | -0.258 | -0.400* |
| 3 | mas | 48.520 | 18.266 | 0.109 | -0.027 | | -0.032 | -0.120 | 0.042 | 0.218 | -0.018 | -0.147 | -0.102 | 0.339 | -0.014 | 0.133 | -0.217 | -0.063 | 0.123 |
| 4 | uai | 66.710 | 24.089 | 0.112 | -0.152 | 0.004 | | -0.234 | 0.171 | -0.304 | -0.122 | -0.257 | -0.08 | -0.271 | 0.452* | -0.608** | -0.066 | 0.022 | 0.022 |
| 5 | ltowvs | 46.632 | 25.141 | 0.005 | 0.000 | -0.089 | -0.111 | | -0.664*** | 0.046 | -0.200 | 0.192 | 0.071 | 0.004 | -0.225 | 0.102 | 0.264 | 0.450* | -0.019 |
| 6 | ivr | 48.449 | 23.302 | -0.291 | 0.245 | 0.005 | -0.078 | -0.533** | | 0.033 | 0.424* | -0.030 | -0.075 | 0.131 | 0.180 | -0.131 | -0.494** | -0.666*** | 0.059 |
| 7 | Social position: People in their 70s | 5.299 | 0.925 | 0.108 | -0.248 | 0.258 | -0.336 | -0.167 | 0.110 | | 0.432* | 0.636*** | 0.332 | 0.535** | -0.003 | 0.378* | -0.042 | -0.143 | 0.172 |
| 8 | People over 70: are seen as friendly | 2.775 | 0.256 | 0.017 | -0.034 | -0.007 | -0.169 | -0.342 | 0.406* | 0.500** | | 0.561** | 0.598** | 0.300 | 0.161 | 0.075 | -0.008 | -0.457* | 0.134 |
| 9 | People over 70: are seen as competent | 2.320 | 0.369 | 0.180 | -0.225 | -0.089 | -0.209 | -0.062 | -0.038 | 0.669*** | 0.610*** | | 0.563** | 0.269 | 0.057 | 0.280 | 0.100 | -0.130 | 0.324 |
| 10 | People over 70: viewed with respect | 2.915 | 0.333 | 0.453* | -0.229 | -0.059 | -0.034 | -0.169 | -0.105 | 0.410* | 0.634*** | 0.639*** | | -0.289 | 0.045 | 0.325 | 0.343 | 0.070 | 0.296 |
| 11 | Is a 70-year old boss acceptable | 6.084 | 0.927 | -0.077 | -0.288 | 0.405* | -0.248 | -0.067 | 0.130 | 0.614*** | 0.323 | 0.326 | -0.141 | | -0.036 | -0.014 | -0.252 | -0.382* | -0.085 |
| 12 | Older people are not respected much these days | 2.775 | 0.217 | 0.126 | -0.305 | 0.006 | 0.326 | -0.191 | 0.219 | 0.057 | 0.168 | 0.059 | 0.032 | 0.025 | | -0.081 | -0.253 | -0.128 | 0.511** |
| 13 | Older people get more than their fair share from the government | 2.039 | 0.386 | 0.299 | -0.272 | 0.218 | -0.574** | 0.066 | 0.010 | 0.477** | 0.111 | 0.267 | 0.253 | 0.219 | 0.015 | | 0.298 | 0.178 | 0.266 |
| 14 | Older people are a burden on society | 1.810 | 0.122 | 0.389* | -0.365* | -0.157 | -0.048 | 0.209 | -0.446* | 0.028 | 0.008 | 0.124 | 0.329 | -0.135 | -0.231 | 0.324 | | 0.473* | -0.014 |
| 15 | Companies that employ young people perform better than those that employ people of different ages | 2.238 | 0.331 | 0.681*** | -0.576** | 0.086 | 0.050 | 0.035 | -0.499** | 0.208 | -0.148 | 0.165 | 0.299 | 0.013 | -0.067 | 0.286 | 0.439* | | 0.167 |
| 16 | Old people have too much political influence | 2.376 | 0.252 | 0.630*** | -0.643*** | 0.223 | 0.059 | -0.223 | -0.022 | 0.387* | 0.246 | 0.458** | 0.435* | 0.194 | 0.416* | 0.345 | 0.080 | 0.517** | |
| 17 | Population ages 65 and above (% of total population) | 11.679 | 5.157 | -0.474** | 0.552** | -0.116 | 0.126 | 0.477** | -0.080 | -0.497** | -0.331 | -0.397* | -0.388* | -0.340 | -0.068 | -0.252 | -0.098 | -0.604*** | -0.548** |
| 18 | Change in percentage aged 65 and above | 1.729 | 1.146 | -0.221 | -0.026 | 0.150 | 0.108 | 0.474** | -0.074 | -0.143 | -0.261 | -0.252 | -0.299 | 0.120 | 0.037 | 0.216 | 0.075 | -0.147 | -0.097 |
| 19 | Log of Gross domestic product per capita, constant prices | 10.098 | 0.683 | -0.441* | 0.494** | -0.140 | -0.266 | 0.321 | 0.281 | -0.242 | -0.176 | -0.338 | -0.387* | -0.212 | 0.054 | -0.001 | -0.093 | -0.570** | -0.537** |

n = 31. *** $p < 0.001$, ** $p < 0.01$, * $p < 0.05$.    PDI: power distance. IDV: individualism. UAI: uncertainty avoidance. MAS: masculinity. LTO: long-term orientation. IVR: indulgence versus restraint.

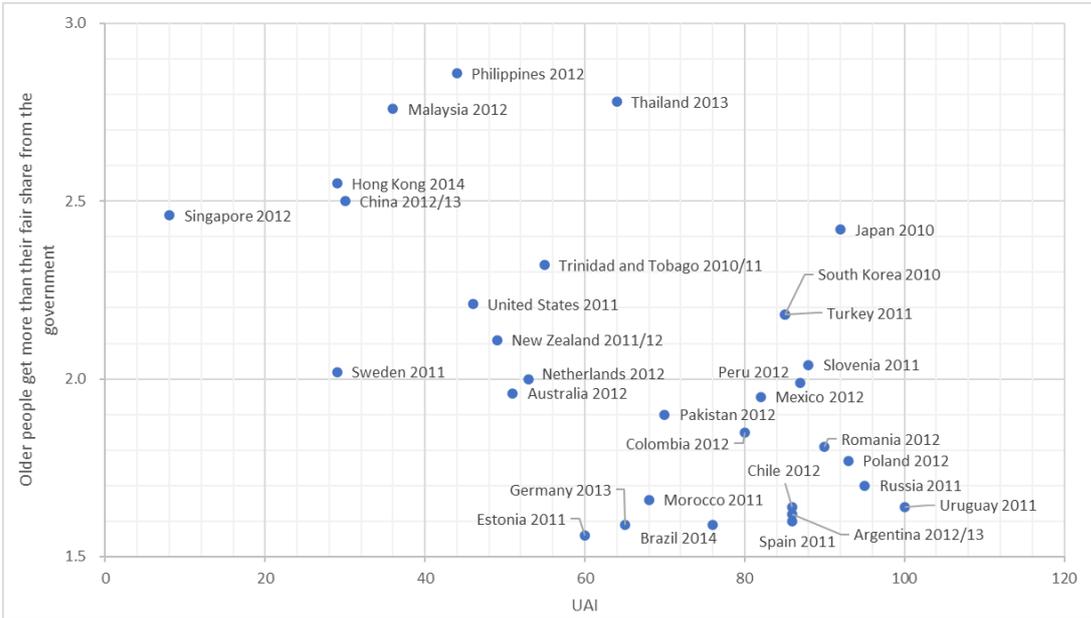

Figure 1. Scatter plot showing the relationship between UAI and "Older people get more than their fair share from the government". UAI: uncertainty avoidance.

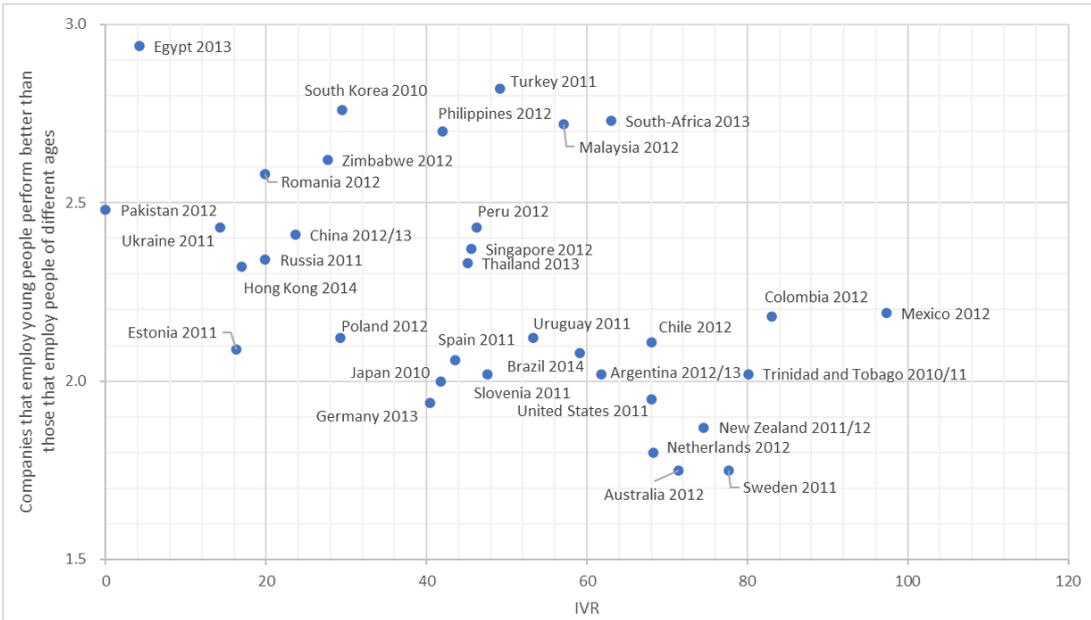

Figure 2. A scatter plot showing the relationship between IVR and "Companies that employ young people perform better than those that employ people of different ages". IVR: indulgence versus restraint.

5. Discussion

    This is the first study to show that Hofstede's various cultural axes are related to various

ageism. Previous studies have shown that IDV [5] is negatively correlated with ageism, and UAI [4,11,13], MAS [14], and TIOWVS [11,14] are positively correlated with ageism, but to the author's knowledge, there has been no research clarifying the relationship between PDI and IVR and ageism.

Of these, previous research has pointed out that PDI overlaps with Confucian societies [8], and it has also been found that Eastern cultures have more negative attitudes toward the elderly than Western cultures [5]. Therefore, a positive relationship between PDI and ageism was predicted, albeit paradoxically. In this study, consistent with these predictions, the higher the PDI, the more the elderly are perceived as a burden to society, inferior to young people in terms of work performance, and too much political influence. On the other hand, the fact that the elderly are respected by society in high PDI societies from the perspective of others (although this became non-significant after the introduction of control variables) may at first glance contradict the results above, but it can be said to be a useful material for interpreting the overall results of this study. In other words, it is shown that people living in countries with high PDI have a strong negative attitude toward the elderly, even though they are aware that they live in a society where the elderly are respected. This is consistent with the paradoxical finding that the suicide rate of the elderly is high in Confucian societies where the elderly are supposed to be protected [10]. Furthermore, considering the geographical overlap between PDI and collectivism [1], this is consistent with previous research that found a positive relationship between collectivism and ageism and argued that collectivist aggression may be directed at older people who do not contribute to society [5]. This study also confirmed that societies with stronger IDV tend to have lower levels of some forms of ageism.

On the other hand, the relationship between IVR and ageism is a topic that has hardly been taken into consideration in previous studies. However, there has been research on aging that has shown that, from the perspective of human development, more fulfilling societies tend to have lower mortality rates [1] and longer life expectancies [17]. This is thought to be because in fulfilling societies, subjective happiness is higher and happiness is perceived positively, which reduces deaths from stress-related diseases such as cardiovascular disease and makes it more likely that elderly people are healthy. In a society where elderly people are lively, it is thought that discrimination and prejudice based on age are less likely to occur. Considering these arguments, it is a convincing result that, as shown in this study, the more fulfilling a society is, the more the perspective of others that the elderly are viewed as friendly by society is shared, and the awareness that the elderly are a burden and the underestimation of their performance is suppressed.

The positive correlation between the lTOWVS and ageism shown in this study is consistent with previous research [11, 14]. Previous research has argued that the reason for the correlation is that investing in young people can be expected to produce greater returns than investing in older people by taking a long-term perspective. Therefore, the correlation between the lTOWVS and the subjective evaluation of the poor job performance of older people compared to younger people shown in this study is quite reasonable. On the other hand, the correlation between the MAS and ageism shown in previous research on English-speaking countries [14] was not shown in this study. The story argued by Ng & Lim-Soh [14] that a society that values competition and highly values the strong and successful is likely to label the elderly as weak, who are the opposite, seems reasonable at first glance, but the results of this study show that this is not necessarily a global trend.

The relationship between UAI and ageism shown in this study requires careful interpretation. Previous studies have found that countries with higher UAI tend to have stronger ageism (Ackerman & Chopik, 2021; Lawrie et al., 2020; Löckenhoff et al., 2009), and have argued that aging is an unpredictable experience as a justification for this [11,12]. However, I find this argument a little difficult to understand. If aging is highly unpredictable, it is not surprising that there is a reverse incentive to prepare an environment that is friendly to the elderly in advance in anticipation of one's own aging future. In this study, people living in countries with strong UAI shared the other-perspective ageism that the elderly are not respected, but were negative about the subjective evaluation that the elderly are taking too much of the government's share. This result suggests that while people are aware of the strong public criticism of the elderly, they also have a strong desire to avoid the confusion that would result from intergenerational redistribution or a reduction in their own future shares. Considering the original meaning of UAI, this result seems easier to understand than previous claims.

6. Limitation

This study analyzed the factors of ageism through cross-sectional analysis. Therefore, it should be noted that the results of this study are correlational and do not indicate a causal relationship. In relation to this, the small sample size also raises concerns about the reproducibility of the results. In addition, because the data was obtained from an existing database, it is possible that the unique cultural differences of each country may have been underestimated. Future research should verify and develop the results of this study through longitudinal analysis or individual analysis using a larger sample. This is the first study to clarify the relationship between Hofstede et al.'s [1] cultural scale and ambiguous ageism.

7. Conclusion

Today, as the aging of the global population accelerates, it is meaningful to clarify the relationship between a country's culture and ageism. In this study, by analyzing data from 40,869 people from 31 countries collected in the World Values Survey Wave 6 and Hofstede's cultural scale, we showed that the five cultural scales of PDI, IDV, UAI, lTOWVS, and IVR are related to ageism after controlling for economic and demographic factors.